\documentclass[journal]{vgtc}                     % final (journal style)
\onlineid{1761}
\vgtccategory{Research}
\vgtcpapertype{application/design study}
\title{%Interactive Visual Design-of-Experiment for\\Optimizing the Configuration of a Cooling System \\ 
Interactive Design-of-Experiments: Optimizing a Cooling System%
}
\author{%
  {Rainer Splechtna},
  \authororcid{Majid Behravan}{0000-0001-6525-6646}, 
  Mario Jelovi\'{c}, 
  \authororcid{Denis Gra\v{c}anin}{0000-0001-6831-2818},  
  \authororcid{Helwig Hauser}{0000-0003-0395-3192}, and 
  \authororcid{Kre\v simir Matkovi\'c}{0000-0001-9406-8943}%
}
\authorfooter{%% insert punctuation at end of each item 
  \item Rainer Splechtna and Kre\v simir Matkovi\'c are with the VRVis Research Center in Vienna, Austria.
  	E-mail: \{Splechtna, Matkovic\}@VRVis.at  
  \item Majid Bahravan and Denis Gra\v{c}anin are with Virginia Tech, Blacksburg, VA, USA.
  	E-mail: \{behravan, gracanin\}@vt.edu  
  \item Mario Jelovi\'{c} is with AVL-AST doo, Zagreb, Croatia.
  	E-mail: mario.jelovic@avl.com.  
  \item Helwig Hauser is with University of Bergen, Norway.
  	eMail:~Helwig.Hauser@UiB.no  
}
\abstract{The optimization of cooling systems is important in many cases, for example for cabin and battery cooling in electric cars.  
Such an optimization is governed by multiple, conflicting objectives and it is performed across a multi-dimensional parameter space. % an automatic solution is not feasible and the user has to approve the compromise solution.
The extent of the parameter space, the complexity of the non-linear model of the system, % high dimensionality and non-linearity of the solution space 
as well as the time needed per simulation run and factors that are not modeled in the simulation necessitate an iterative, semi-automatic approach. % to finding a desirable solution.  
We present an interactive visual optimization approach, where the user works with a p-h diagram to steer an iterative, guided optimization process.
A deep learning (DL) model provides estimates for parameters, given a target characterization of the system, while numerical simulation is used to compute system characteristics for an ensemble of parameter sets.  
Since the DL model only serves as an approximation of the inverse of the cooling system and since target characteristics can be chosen according to different, competing objectives, an iterative optimization process is realized, developing multiple sets of intermediate solutions, which are visually related to each other.   
The standard p-h diagram, integrated interactively in this approach, is complemented by a dual, also interactive visual representation of additional expressive measures representing the system characteristics.  
We show how the known four-points semantic of the p-h diagram meaningfully transfers to the dual data representation. 
When evaluating this approach in the automotive domain, we found that our solution helped with the overall comprehension of the cooling system and that it lead to a faster convergence during optimization.  
}
\teaser{%
  \centering
  \vspace{2ex}
  \includegraphics[width=\linewidth]{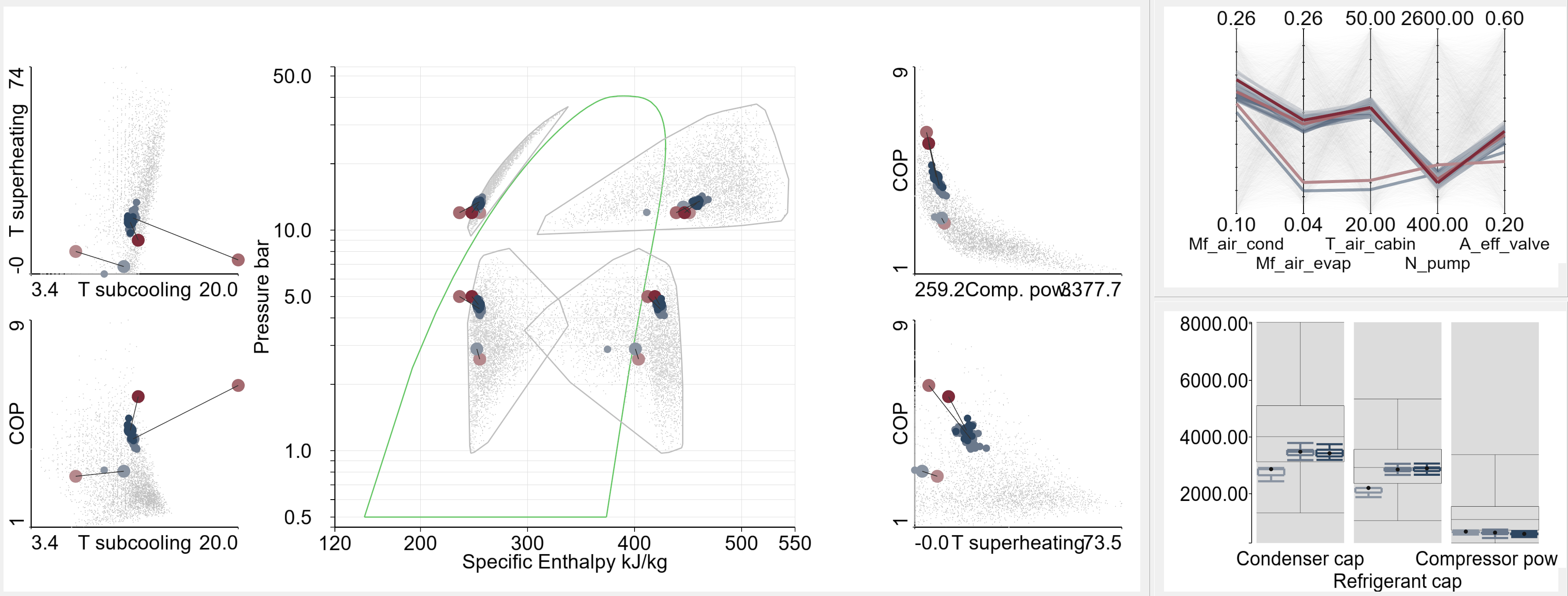}
  \caption{%
  Key views of the visualization component in iterative design of experiments. The interactive p-h diagram, central to cooling system design, presents multiple layers of information: user-defined desired points (in shades of red), simulated points generated by parameters predicted through deep learning (shades of blue), and scatterplots offering a dual data perspective (with lines connecting Deep Learning prediction and simulation for the same parameters). Parallel coordinates show control parameters, while box plots offer insights into numerical aggregated values. The colors in all views are consistent. Darker colors show recent iterations.
  \\~ 
  }
  \label{fig:teaser}
}
\graphicspath{{Figures/}{figures/}{pictures/}{images/}{./}} % where to search for the images
\usepackage{tabu}                      % only used for the table example
\usepackage{booktabs}                  % only used for the table example
\usepackage{lipsum}                    % used to generate placeholder text
\usepackage{mwe}                       % used to generate placeholder figures
\usepackage{mathptmx}                  % use matching math font
\begin{document}
%
%%%%%%%%%%%%%%%%%%%%%%%%%%%%%%%%%%%%%%%%%%%%%%%%%%%%%%%%%%%%%%%%
%%%%%%%%%%%%%%%%%%%%%% START OF THE PAPER %%%%%%%%%%%%%%%%%%%%%%
%%%%%%%%%%%%%%%%%%%%%%%%%%%%%%%%%%%%%%%%%%%%%%%%%%%%%%%%%%%%%%%%
%
%% The ``\maketitle'' command must be the first command after the
%% ``\begin{document}'' command. It prepares and prints the title block.
%% the only exception to this rule is the \firstsection command
\firstsection{Introduction\label{sec:introduction}}

\maketitle

In today's world, the importance of cooling systems has surged notably owing to the urgent imperative to curtail $\mathrm{CO}_2$ emissions and to pioneer more efficient housing and transportation.
Electric vehicles are an important example, where the necessity for cooling extends beyond passenger comfort to also ensure the crucial cooling of batteries. 
%The intricate and nonlinear nature of cooling systems necessitates an interactive and exploratory approach to their design.

Several factors exacerbate the challenge of optimizing the design of a modern cooling systems.  
The optimization has to deal with \textit{\textbf{multi-dimensional control and output parameter spaces}} of the simulation models of such thermodynamic systems.
These models are characterized by their \textbf{\textit{non-linear behavior}} and can only be linearized locally with a small range of sufficient validity. 
To be close to the physical reality, the simulation models are usually formulated as systems of ordinary differential equations or as differential algebraic equation systems. %  herefore the physical system models, instead of black box models, are often solved numerically in form of an ordinary differential equation system (ODE) or differential algebraic equation system (DAE). 
Due to the computational complexity of the necessary iterative numerical simulation, it is not feasible to densely sample the multi-dimensional parameter space % The solving of a physical model of such a system is far more complex than solving a black box model 
and \textbf{\textit{inverting the simulation model is not possible}} to derive the system's configuration directly from a certain, desired behavior.  
% Furthermore the simulation model is also an approximation and, though "the best one can have in silico", too costly to use on a massive scale, e.g., for dense sampling of the control parameters.
Accordingly, a \textbf{\textit{design of experiments approach is needed}} to optimize across the multi-dimensional parameter space effectively and efficiently. %  sample the parameter space sparsely.
This methodology is commonly applied to simulations utilized in the automotive domain, and also in the use case presented in this paper. It encompasses planning, conducting, analyzing, and interpreting controlled simulation runs with the goal of gaining insight into a system's behavior with a minimal set of cases, i.e., simulation runs, by applying statistical methods and, typically, varying multiple parameters simultaneously.
Parameter sweeps are also used to explore parameter spaces by issuing simulation runs with varying parameters per case, but we adopt the term design of experiments due to its broader scope and because it is established in the automotive domain.

It should also be possible to
\textbf{\textit{consider context knowledge of the engineers}}, for example, regarding the expected, limited accuracy of the simulation model, as well as other external factors such as the availability of system parts, safety concerns, policies, etc.

Identifying suitable control parameters that reliably lead to a certain, optimized behavior of the cooling system under a variety of possible operating conditions, while also respecting expectations with respect to other related aspects (costs, safety, policies, etc.), amounts to a complex \textbf{\textit{exploratory task}}.  Supporting this task with a mixed-initiative solution, empowering the expert user while helping with computational optimization, based on numerical simulation as well as on machine learning, promises to limit the number of necessary exploratory iterations.  
%
% Consequently, identifying the control parameters for achieving the desired outputs, while also incorporating the knowledge of the engineers and trying to keep constraints under control (that the explored system variants remain physically possible, mechanically doable, economically reasonable, within the realm of policies, etc.) becomes a laborious and error-prone \textbf{\textit{exploratory, iterative task}}.

%In order to comprehend the system behavior 

%* different space that need to be linked, including, at least, the space of simulation parameters, and the space of [some] simulation output

%* context knowledge of the engineers, for ex., regarding the expected accuracy of the simulation model (or the lack thereof, rather), other, external factors such as availability of system parts, safty concerns, policies, etc.

%* the simulation model (which also is an approximation), which is "the best one can have in silico", though -- but it's too costly to use it on a massive scale, for ex., for dense sampling of parameters, etc.

%* exploratory design by the engineers, trying to find better system configurations, while also trying to keep constraints under control (that the explored system variants remain physically possible, mechanically doable, economically reasonable, within the realm of policies, etc.)

%Th(Interactive DoE!) that integrates different components to enable the needed iterated optimization workflow.
%

\begin{figure}[t!]
\centering
    \includegraphics[width=.8\linewidth]{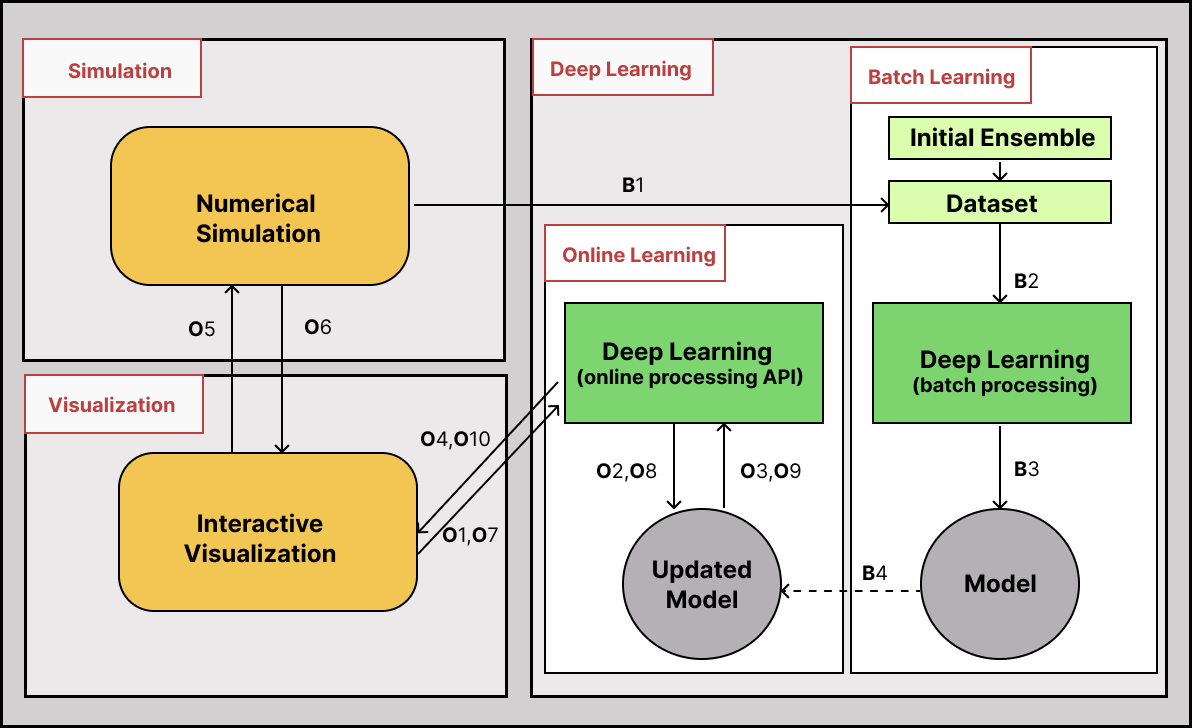}
    \caption{The IDoE approach. Labels prefixed with a `B' relate to batch learning, those with an `O' correspond to online learning.}
    \label{fig:chartDLModel}
\end{figure}

In the following, we present our approach to the interactive visual exploration and optimization of a complex systems, such as described above, that we call ``Interactive Design of Experiments'' (IDoE).
This IDoE approach integrates the following components (\autoref{fig:chartDLModel}) to facilitate an improved, iterative, exploratory optimization workflow:
%that focuses particularly on regions likely to produce desired outcomes:
%
\begin{enumerate}
\item \textbf{Initial ensemble}: an early offline component, providing a starting sampling of the parameter space of the simulation (costly and still sparse, even though providing quite a number of simulation runs -- sparse, because the parameter space is multi-dimensional).
\item \textbf{Interactive visualization}: provides guided interaction to facilitate the expert's exploration of possible improvements of the system configuration and establishes a visual relation between parameters and the simulated system behavior. % specification of desired output characteristics.
\item \textbf{Deep learning (DL)}: an approximate inversion model provides rapid responses for desired output parameters, estimating reasonable control parameters for regions of interest in the parameter space, specified interactively in the visualization. % The model is trained with data from the initial ensemble.
%some "re-fitting" to doable configurations, bringing the exploration back to what's possible 
%
\item \textbf{Numerical simulation}: provides iterative refinements of the initial ensemble based on regions of interest in the parameter space as explored iteratively in the visualization; new simulation runs are also used to further improve the DL model. %The additional data are also used for refining/adapting the DL model on the fly via online/incremental learning.
%(doing a couple of computational, automatic optimization steps)  
\end{enumerate}
\noindent 
The visualization component serves as the central interface, enabling the user to interact with and to interpret the data.
We exploit an initial ensemble, computed offline, and actively guide the refinement process by specifying desired outcomes in the interactive visualization.
Using our DL inversion model, we estimate 
% Utilizing a DL model, we calculate 
the corresponding control parameters for these outcomes.
Numerical simulation then computes the actual outcomes for these estimated parameters.  
% We validate the proposed solution and its vicinity in the parameter space through simulation.
This guided, iterative process enables to progressively improve the sampling of the parameter space, focusing particularly on regions that likely produce desired outcomes, hence the term \textbf{\textit{Interactive Design of Experiments}} (IDoE).
This iterative process is usually structured by the multiple objectives or scenarios the system should be optimized for.
Each scenario is individually explored through several iterations.

%\medskip\par 
For our use case, i.e., the optimization of air conditioning (AC) systems, the main view in the visualization component is an interactive p-h diagram (\autoref{fig:ph diagram general}), which illustrates the relation between pressure and specific enthalpy of the refrigerant, corresponding to a pivotal standard diagram in AC system design.  
In our interactive realization of this diagram, we integrate guided interaction to aid the specification of desired output parameters.
We also extend it by four scatterplots---one related to each characteristic point---realizing an informative, dual perspective on the data.

The system's design is the result of a collaboration with domain experts, one of whom is also a coauthor of this paper. We present a practical use case to demonstrate the effectiveness of our approach. The IDoE approach, however, likely has a much broader applicability to optimizing complex simulation models that cannot be inverted, i.e., a very common case.
The main contributions of this application paper can be summarized as follows:
\begin{enumerate}
    \item The Interactive Design of Experiments approach, enabling a rapid, guided optimization across a multi-dimensional parameter space, based on an initial simulation ensemble that is step-wise improved using two models---one for the numerical simulation of the system, as well as one for the approximate inversion of it. % , as well as which allows for the initiation with a relatively small simulation ensemble, followed by iterative refinement in an interactive manner. 
%    This method ensures that additional runs are strategically computed in areas likely to yield desired outputs, particularly beneficial for simulation models that cannot be easily inverted.
%
    \item Our visualization design of an interactive p-h diagram, enriched by a dual perspective on the data, and 
    % A highly configurable, interactive p-h diagram, 
    serving as the central interface for the guided, interactive optimization.
    \item A use case, demonstrating the successful application of IDoE in tuning a cooling system in the automotive domain.
    The simulation model of the system in this use case has five control parameters, the initial ensemble consists of 5000 runs and a single run takes about ten seconds. 
    Usually the analysis includes about ten scenarios, like "hot summer day", with up to a dozen iterations each.
    We showcase the exploration of one scenario.

%    simulation model, along with the utilization of the interactive p-h diagram.
%   
\end{enumerate}

%%%%%%%%%%%%%%%%%%%%%%%%%%%%%%%%%%%%%%%%
%%%%%%
%%%%%%  BACKGROUNMD
%%%%%%
%%%%%%%%%%%%%%%%%%%%%%%%%%%%%%%%%%%%%%%%

\section{Background }
\label{sec:background}
\begin{figure}[t!]
\centering
\includegraphics[width=.8\linewidth]{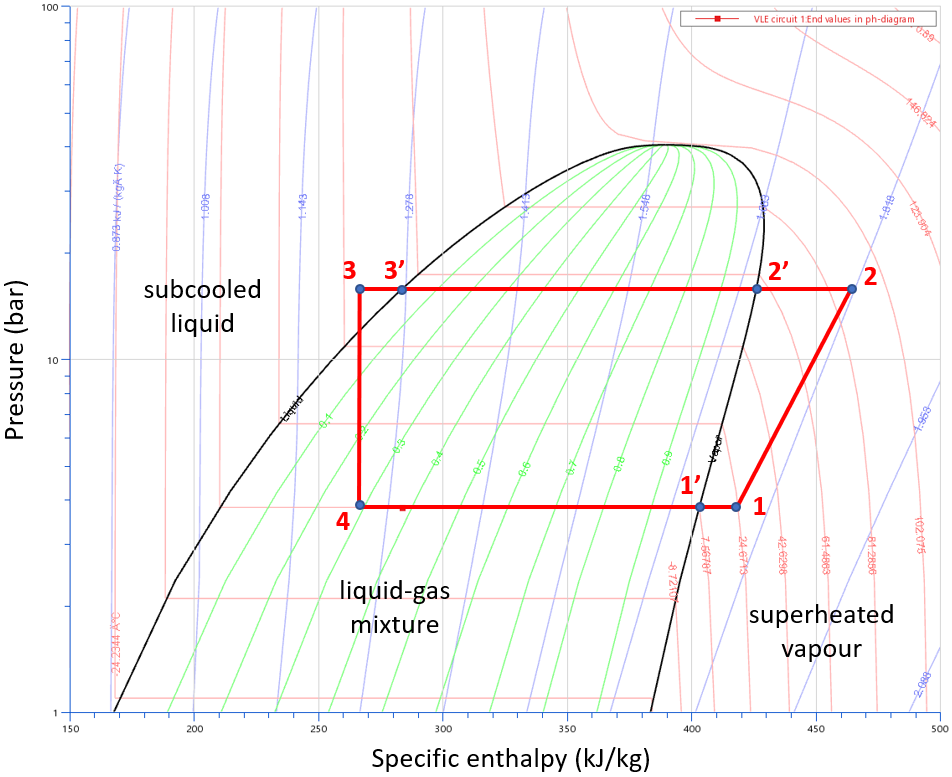}
\caption{An example of a p-h diagram showing the refrigeration cycle in counterclockwise direction. The characteristic points are shown in red. The connecting lines between the points correspond with state changes in the cycle:  Compression (1$\rightarrow$2), condensation (2$\rightarrow$3), expansion (3$\rightarrow$4) and evaporation (4$\rightarrow$1). Condensation and evaporation pressures can also be easily identified in the diagram.}
\label{fig:ph diagram general}
\end{figure}
For decades, automotive air conditioning (AC) systems were primarily designed to cool the vehicle's interior. 
Their significance has recently surged, especially with the rise of electric vehicles, where AC systems play a crucial role in cooling batteries, inverters, and electric motors. 
The complexity of these systems demands meticulous thermal solutions, and their development process is time-intensive. 
Inefficient utilization can lead to substantial costs.
Hence, it is imperative to analyze and predict the behavior of AC systems
%across different operating modes
at the vehicle design stage. 
The early consideration of involved factors is essential for optimizing performance and minimizing potential drawbacks.

All AC systems operate according to fundamental thermodynamic principles, employing similar components and utilizing a refrigerant to achieve the desired cooling effect. The refrigerant undergoes temperature and pressure changes through the action of a compressor, transitioning between liquid and gaseous states in a closed system. 
% , isolated from the external environment.

Typically, an AC system consists of four main components: the compressor, the condenser, an expansion valve, and the evaporator. While each component serves a specific function, they collaborate to realize the AC's functionality. 
The compressor pressurizes the gaseous refrigerant, directing it to the condenser, where heat is expelled, causing the gas to condense into a liquid under high pressure. 
Subsequently, the refrigerant passes through the expansion valve, where its pressure decreases, before entering the evaporator. 
Within the evaporator, the refrigerant evaporates, absorbing heat from the surrounding air (or another medium in the case of chillers), thereby cooling it. 
This heat exchange process facilitates the refrigerant's return to its gaseous state, and the cycle repeats.

Engineers represent this cycle in a pressure-enthalpy (p-h) diagram, which shows the thermodynamic properties of the refrigerant in the cooling system (\autoref{fig:ph diagram general}). The p-h diagram shows the refrigerant's pressure on the y-axis (logarithmic scale) and its specific enthalpy on the x-axis (linear scale).
It provides a visual representation of the refrigerant's thermodynamic state as it traverses through the system's components throughout the cooling cycle.
%The cycle operates counterclockwise on the p-h diagram, with changes of state occurring at each of the four main components: compressor, condenser, expansion valve, and evaporator.
The refrigeration cycle shows up in a counterclockwise direction in the p-h diagram, with changes of state occurring at each of the four main components of the AC circuit:
\begin{enumerate}
\item \textbf{Compression} (1$\rightarrow$2) of the gaseous refrigerant to condensing pressure at the compressor component. 
\item \textit{Vapor cooling} (2$\rightarrow$2'), \textbf{condensation} (2'$\rightarrow$3'), and \textit{subcooling} (3'$\rightarrow$3) of the liquid at the condenser component. 
\item \textbf{Expansion} (3$\rightarrow$4) of the liquid-gas mixture to evaporation pressure at the expansion valve component.
\item \textbf{Evaporation} (4$\rightarrow$1') and \textit{superheating} (1'$\rightarrow$1) of the vapor at the evaporator component.
\end{enumerate}
Points 1, 2, 3, and 4 in the  p-h diagram (\autoref{fig:ph diagram general}) are indicators of the efficiency of the cooling cycle and are called characteristic points.

Typically, when designing a cooling system the engineers do not specify these points directly when describing the desired cooling cycle characteristics.
They specify low (evaporating) and high (condensing) pressures and subcooling and superheating temperatures of the cycle instead.
The characteristic points can be calculated from these values.
%Additionally, specific properties of the refrigerant, including temperature, specific volume, specific entropy, and gas content, can be derived from the pressure-specific enthalpy diagram.
%

%%%%%%%%%%%%%%%%%%%%%%%%%%%%%%%%%%%%%%%%
%%%%%%
%%%%%%  Related Work
%%%%%%
%%%%%%%%%%%%%%%%%%%%%%%%%%%%%%%%%%%%%%%%

\begin{figure*}[t!]
\includegraphics[width=\linewidth]{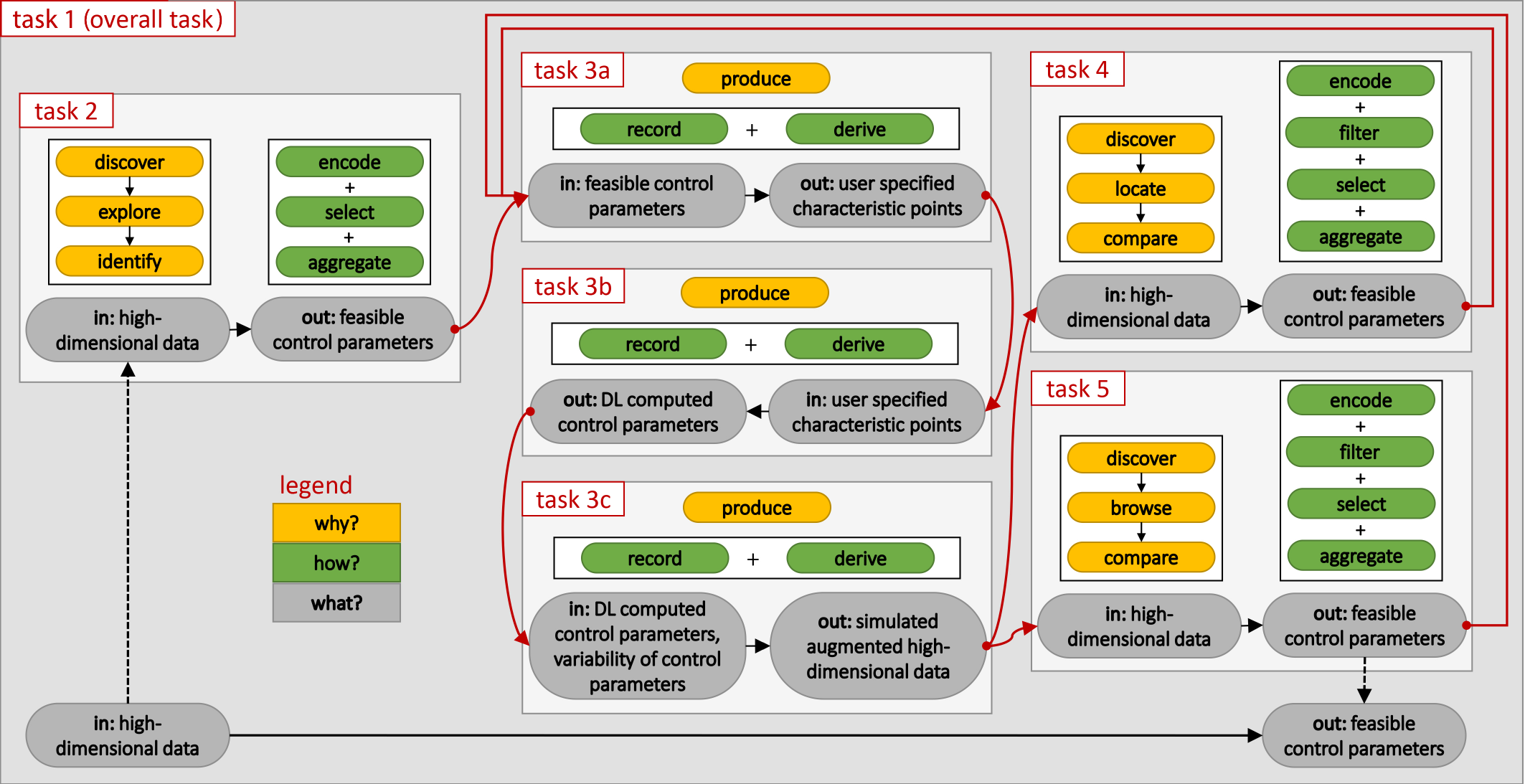}
\caption{Sequence of interdependent tasks in the IDoE workflow.}
\label{fig:tasks}
\end{figure*}

\section{Related Work}

%Related Work is still missing

AC systems consist of many components, including compressor, condenser, evaporator, tube, and thermostatic expansion valve. %~\cite{Alawadhi-2022-a}.
Hence, to simulate an AC system, a number of control parameters are used to specify its characteristics.
The AC design space is very large and would benefit from approaches similar to ones shown for other domains, like the use of computational steering as shown by Atanasov et al.~\cite{Atanasov-2010-a} or Mulder et al.~\cite{Mulder-1999} and parameter space exploration as described by Sedlmair et al.~\cite{Sedlmair-2014-a} to identify desirable system configurations.

Specifically, the functioning of AC systems heavily depends on using a refrigerant to exchange heat and produce a cooling effect.
Knowing the thermophysical properties of refrigerants, including the relationship between pressure and enthalpy, is essential to identify the values of control parameters used to design and optimize AC systems.
There are many software libraries that support high-accuracy simulation and evaluation of the thermophysical properties of fluids like CoolProp~\cite{CoolPropLibrary}.
That also includes a number of machine learning based approaches as described by Abdul Jameel et al.~\cite{inproceedings}, Folmsbee et al.~\cite{doi:10.1021/acs.jpca.0c10147} and Waxenegger-Wilfing~\cite{pr10112384}.

Thermophysical properties  can be visualized using various types of thermodynamic state diagrams, including
pressure-enthalpy (ph),
pressure-specific volume (pv),
pressure-temperature (pT),
temperature-enthalpy (Th), and
temperature-entropy (Ts).

The p-h diagram (\autoref{fig:ph diagram general}) is an important graphical representation that shows the relationship of pressure and specific enthalpy. It is used frequently to determine the performance of AC systems and show how thermodynamic processes are working with changes of states as described by Aly et al.~\cite{Aly-1981-a}.
Pressure is shown on the y-axis (logarithmic scale) and enthalpy on the x-axis (linear scale) as described by Khennich et al.~\cite{Khennich-2014-a}.
The p-h diagram shows how pressure and specific enthalpy relate to each other at certain boundary conditions.

There are many online and standalone general visualization tools, like MATLAB ~\cite{MATLAB}, maple ~\cite{maple} or Mathematica~\cite{Mathematica} and specialized visualization tools, like NIST~\cite{CoolSelector} or CoolSelector ~\cite{CoolSelector} that support the exploration of p-h diagrams.
Our approach differs, as it enables the user to specify wanted points and provides the corresponding parameters then.
We also offer a dual perspective on the data.

Recently, DL methods have been used to analyze and control cooling systems as presented by Chen et al.~\cite{Chen-2019-a}, Chervonyi er al.~\cite{Chervonyi-2022-a} and Luo et al.~\cite{Luo-2022-a}.
In the exploration of Deep Neural Networks (DNNs) within an online setting, there are significant challenges and solutions, reflecting scenarios where data is not sequentially available, as opposed to traditional batch settings where DNNs are trained with the entire training data accessible from the outset, a method that becomes untenable in many real-world situations as shown by Sahoo et al.~\cite{Sahoo2017Online}.
This is particularly true when data continuously accumulates over time or when its volume exceeds memory storage capabilities.
This distinction underscores the necessity for adaptive approaches that accommodate the ongoing arrival of new information, thereby necessitating the exploration of the parameter space, its segmentation, and advanced techniques for interactive visual analysis in simulation environments.

Numerical simulations can produce extensive, multi-dimensional datasets, posing challenges for exploration.
Innovative visualization approaches, like Bonneau et al.~\cite{Bonneau-2003-a}, are necessary to navigate and understand such complex data effectively.
Typically, interactive visual analysis of simulation data employs customized views to illuminate various facets of multi-dimensional and spatio-temporal data as presented by Kehrer and Hauser~\cite{KH2012}.
However, not all visualization techniques, tools, and views are optimal for studying simulation data.
Selecting suitable visualizations, tailored to the specific simulation domain and data characteristics, can benefit from a taxonomy of visualization solutions. Vernon-Bido et al. ~\cite{Vernon-Bido-2015-a} outline such a taxonomy within the realm of modeling and simulation.
In our case, we introduce the novel, interactive p-h view, integrating advanced visualization features while retaining the engineers' familiarity with their preferred view.

P\'{a}lenik et al.~\cite{IsoTrotter} introduce a mixed-initiative approach to visually guided modeling for atmospheric convection, integrating visual parameter space analysis with partial automatic parameter optimization.
Although it relates to a different domain, the similarity of the IsoTrotter and our approach lies in depicting data in dual spaces.
We use aggregated and computed data in addition to the p-h diagram, and they use the parameter space to constrain sampling. 

Finally, steering of scientific simulations can be realized with a surrogate model as described by Bonneau et al.~\cite{Bonneau-2003-a} or without a surrogate model as shown by Matkovi\'{c} et al.~\cite{steering}.
Berger et al.~\cite{Berger} describe a system for conducting interactive, prediction-based local analysis of parameter spaces using methods from statistical learning.
Splechtna et al.~\cite{Splechtna} use the output parameters of models of the same system at different complexity levels to map the different control parameters of these models and enable brushing and linking across models.

In our case we use a surrogate (DL) model to inverse the simulation, not to replace it, i.e., our approach is output parameter centric, in the sense, that in contrast to the aforementioned works, the start point or focal point~\cite{Berger} is not set in the control parameter space but in the output parameter space.
In this way we can predict control parameters for the desired outputs of a given scenario, and steer the ensemble creation efficiently.
Most approaches use similar visualization techniques, like coordinated multiple views with linking and brushing, and views, like scatter plots or parallel coordinates.
We augment these techniques and views to highlight the data important for the current task at hand.

%AC simulation/design/visualization
%steering computation
%parameter space exploration

%%%%%%%%%%%%%%%%%%%%%%%%%%%%%%%%%%%%%%%%
%%%%%%
%%%%%%  Tasks and Requirements Analysis
%%%%%%
%%%%%%%%%%%%%%%%%%%%%%%%%%%%%%%%%%%%%%%%

\section{Tasks and Requirements Analysis}  \label{section:tasks}
Designing an AC system, like many others, involves various constraints. 
Firstly, there are physical limitations: Depending on the selected refrigerant, only certain cooling cycles can be implemented. 
Additionally, manufacturing constraints, such as the maximum revolving speed of a compressor, come into play. 
If required component characteristics are unavailable due to manufacturing constraints, the system cannot be realized, neither. 
Furthermore, there are simulation constraints: The used simulation model usually only covers a subset of the actual physical system with sufficient accuracy. 

The simulation model of the AC system is based on multiple control parameters that can be chosen by the user.
These control parameters, together with other fixed values, determine the behavior of the system.
The output parameters are then calculated by the simulation for the specified system.
In the context of AC systems, a set of output parameters~-- called characteristic points~-- is of special interest since they are the essential indicators of the efficiency of the AC cycle (\autoref{fig:ph diagram general}).  

The central challenge of cooling system design lies in identifying control parameters  that yield a desired, optimized system behavior, with a special focus on the characteristic points, while also adhering to all other constraints.
%If an inverse model of the simulation were available, the task would be simplified. However, such models are typically unavailable, necessitating extensive experience and expertise from engineers to tune complex systems like AC units.
%Hence when trying to optimize the system we are looking for a combination of control parameters that produce the desired output parameters with focus on the characteristic points.
In principle, a system designer may wish to ``go backwards'', deriving suitable control parameters from the specification of a certain, desired system behavior for a variety of scenarios.
A scenario can be a specific operating condition, like "hot summer day", "Overheated cabin, cold motor start" or another design goal, like "Improve coefficient of performance".
Each scenario is explored and optimized in an iterative fashion.
To inform our design, we engaged in a six-months collaboration of visualization and engineering experts.
Given the established prevalence of the p-h diagram in cooling system design, we considered it necessary to base our visualization design on this standard.
Based on numerous meetings and discussions, we introduced the IDoE approach described in ~\autoref{sec:introduction}
% From the very beginning it was clear that we will use the p-h diagram, as it is so central and common in the AC systems design, analysis, and description and shows these characteristic points.
%To cope with its complexity and variety of information it provides, we made it interactive.
and identified the following tasks.
Following the approach by Brehmer and Munzner~\cite{Brehmer-2013}, we then analyzed these tasks in order to derive crucial requirements for our visualization design.  
% The workflow can be broken down to the following analysis tasks that then inform the requirements described below.
% The described workflow and tasks are the result of a six-month collaboration of visualization and engineering experts and numerous meetings and discussions.
% The task abstraction uses the typology proposed by Brehmer and Munzner~\cite{Brehmer-2013}.
% 
%And then our data and simulation model description, in total section should be 1 page at most
% 
\begin{enumerate}[start=1,label={\bfseries T\arabic*}]
%\begin{enumerate}
    \item Overall task: Comprehend and optimize the AC system, i.e., identify control parameters so that the system operates as efficient as possible for all relevant scenarios.
    \item Explore the p-h diagram, together with its dual representation, the derived values space, and the control parameters space, to maintain a holistic understanding of the system design.
    \item Iteration task: Identify necessary adaptations of the control parameters for an ensemble refinement based on desired cooling cycle characteristics for a specific scenario.
    \begin{enumerate}[start=1,label={\bfseries T3\alph*}]
%    \item Interactively specify desired characteristic points.
    \item Interactively specify desired cooling cycle characteristics.
    \item Estimate related control parameters through an attempted model inversion.
    \item Issue and integrate additional simulation results based on the inferred control parameters.
    \end{enumerate}
    \item Evaluate the ensemble refinements with respect to a specific scenario. %design goals and constraints.
    \item Compare selected iterations of ensemble refinements with respect to all scenarios.
\end{enumerate}
\medskip\par\noindent
Analyzing the above outlined tasks and sub-tasks, we identified the following design requirements:
% The following requirements were identified to support the described tasks:
\begin{enumerate}[start=1,label={\bfseries R\arabic*}]
    \item Depict input and output spaces (control parameters and system outcomes) using linked views and support guided interaction.
    \item Use the standard p-h diagram, as known to the experts, to show and specify cooling cycle characteristics.
    \item Integrate a dual representation of characteristic points~-- keep the arrangement, but show additional, dual data.
    \item Based on the specified characteristic points, suggest related control parameters by attempting a model inversion in order to seed a new ensemble refinement iteration.
    \item Initiate a new ensemble refinement (additional simulation runs) in the vicinity of the suggested control parameters.
    \item Visualize all ensemble refinements and keep track of all iterations.
    \item Provide quantitative data for the iterations and for the individual runs to enable a comparison. 
\end{enumerate}

%%%%%%%%%%%%%%%%%%%%%%%%%%%%%%%%%%%%%%%%
%%%%%%
%%%%%%  Interactive Visual Design of Experiments
%%%%%%
%%%%%%%%%%%%%%%%%%%%%%%%%%%%%%%%%%%%%%%%

\section{Interactive Design of Experiments}
%
%Designing an AC system, like many others, involves various constraints. Firstly, there are physical limitations; depending on the selected coolant, only certain cooling cycles can be implemented. Additionally, manufacturing constraints, such as the maximum revolving speed of a compressor, come into play. If required component characteristics are unavailable due to manufacturing constraints, the system cannot be realized. Lastly, there are simulation constraints; simulations can only model a subset of the real world with limited certainty. The main challenge lies in finding control parameters of the system that yield the desired system behavior while adhering to all constraints. If an inverse model of the simulation were available, the task would be simplified. However, such models are typically unavailable, necessitating extensive experience and expertise from engineers to tune complex systems like AC units.
%
%Our approach provides engineers with support in navigating this complex and constrained space.
Our interactive design of experiments approach needs the following three main components to enable the engineer to perform the tasks as outlined in \autoref{section:tasks}, satisfying the requirements as imposed by the tasks:
\textit{\textbf{the simulation component}}, which simulates the AC system; 
\textit{\textbf{the deep learning component}}, which attempts to invert the simulation model; and 
\textit{\textbf{the interactive visualization component}}, which serves as the central hub for visualization and interaction, as well as the main interface to all components.
\autoref{fig:chartDLModel} illustrates these three overall components % necessary for realizing the interactive design of experiment approach 
and the data exchange between them. 
The IDoE workflow is based on an offline setup phase, followed by the iterative mixed-initiative optimization phase.
%
%with the following steps:
\smallskip\par\noindent
\textbf{Setup phase} (offline, not interactive, once):    
\begin{enumerate}
\item Create an initial ensemble using the simulation model, providing a sparse, yet as-large-as-feasible sample of the parameter space, while considering manufacturing constraints. % This involves sampling the parameter space sufficiently given the available time and computation resources.
\item Train a DL model to attempt an inversion of the simulation model, using the initial ensemble, suggesting likely responsible control parameters for a certain, desired system behavior. % This enables the specification of desired simulation outputs and provides predictions on corresponding control parameters.
\end{enumerate}
\smallskip\par\noindent
\textbf{Mixed-initiative optimization phase} (interactive, iterative~-- see \autoref{fig:tasks} for our corresponding task analysis):    
% \autoref{fig:tasks} shows the interactive DoE workflow:
\begin{enumerate}
\item Visualize and explore the initial ensemble (T2).
\item Specify desired outputs, taking into account physical constraints; in our case, the user specifies four values to define the properties of a cooling cycle in the p-h diagram (T3a).
\item Estimate the control parameters for the specified outputs using the DL model (T3b).
\item Run the simulation for values at and near the estimated control parameters (T3c).
\item Compare the computed and predicted outputs, and sample the parameter space in the vicinity of the predicted parameters to assess the quality of the predicted solution (T4) and/or compare computed and predicted outputs of previous iterations (T5).
%\item Use the simulation results to update the DL model in order to improve its predictions.
\item Iterate or conclude the exploration and export selected cases (T1).
\end{enumerate}

\subsection{The Interactive Visualization Component}

% \begin{figure*}[t!]
%     \centering
%     \includegraphics[width=\linewidth]{Figures/PH1.png}
%     \caption{missing}
%     \label{fig:PH1}
% \end{figure*}

% \begin{figure}[t!]
%     \centering
%     \includegraphics[width=\linewidth]{Figures/Layout.png}
%     \caption{missing}
%     \label{fig:dualSpace}
% \end{figure}

% \begin{figure}[t!]
%     \centering
%     \includegraphics[width=\linewidth]{Figures/BoxPlots1.png}
%     \caption{Will be replaced with a box plot with several iterations}
%     \label{fig:boxplots1}
% \end{figure}

% \begin{figure}[t!]
%     \centering
%     \includegraphics[width=.6\linewidth]{Figures/GUIelement.png}
%     \caption{Will be replaced with a box plot with several iterations}
%     \label{fig:GUIelement}
% \end{figure}

% As the deep learning model cannot fully capture the complexity of the simulation model, human involvement becomes essential. To comprehend and fine-tune the AC system, engineers need to observe results, compare them with the simulation outcomes, and experiment with various characteristic points.

The data we are working with exhibits a non-trivial structure. For a set of scalar control parameters, the simulation solver computes four points in the p-h diagram, along with several other scalar outputs. These four points are interconnected, forming a cycle within the diagram. Additionally, there are supplementary values, such as the coefficient of performance, which can be derived from the four points. These values offer a dual perspective on the cycle and represent a valuable source of information for engineers.

Given the complexity of both the data and the requirements, we have chosen to utilize the well-established Coordinated Multiple Views (CMV) paradigm (R1). This approach has demonstrated effectiveness in ensemble simulation analysis (as well as in many other applications). Attempting to address all requirements within a single integrated view would result in excessive complexity and confusion.

Our design revolves around the p-h diagram. This diagram holds paramount importance in AC system design, making its omission unacceptable. We have enhanced its usability by incorporating interactive features for ease of analysis. In order to display additional data, we augment it with four scatterplots. Each of them is strategically placed to convey information related to a certain point of the cycle. We also utilize parallel coordinates for visualizing control parameters and box plots for presenting quantitative data and facilitating comparisons. Special emphasis has been placed on interaction.

\subsubsection{Interactive p-h Diagram}

The p-h diagram illustrates the characteristic points of the AC cycle through an orthogonal plot, with the logarithm of pressure ($P$) represented on the y-axis and specific enthalpy values on the x-axis (linear scale). Alongside, it displays the saturation curve, iso-quality lines, iso-thermal lines, and iso-entropy lines.
We utilize the p-h diagram to represent characteristic points of the initial simulation ensemble, specify desired cooling cycle characteristics, and depict both specified/calculated and simulated characteristic points (R2). Optionally, connecting quadrilaterals for these points, i.e., the cooling cycle, may also be included, as well as the convex hull of the initial points, which indicates the area resulting from plausible parameters.

While this diagram conveys a wealth of information, it risks becoming cluttered when all elements are displayed simultaneously. Clearly, there is a need for an efficient method to filter out data that is not currently the focus.
For example when specifying the desired cooling cycle characteristics, the details of the initial points are not of interest but the convex hull of the initial points is, since it gives a good indication if the currently specified cooling cycle characteristics, that are represented by the four characteristic points, can be achieved within the system's plausible parameter range (see \autoref{fig:points}a).
%Hence the user would hide/filter the initial points in the p-h diagram and activate the convex hull representation.

% The interactive p-h diagram can display all initial runs (\autoref{fig:PH1}a). These initial runs serve both as training data for the model and as a means to gain better insight into the characteristics of the cooling system.
% During the tuning phase, engineers typically focus on the convex hull of the initial points. In our case, these initial points were generated by sampling the entire plausible space of control parameters, allowing the convex hull to approximate potential areas of characteristic points.
% finally, the desired points defined by the user, as well as the simulated points for the predicted parameters and points in the vicinity should be shown.

In the following, we emphasize the most noteworthy design considerations incorporated during the creation of the interactive p-h diagram.

\paragraph{Guided Points Specification}
Since the four characteristic points of a cooling cycle are interdependent, they are typically indirectly specified by specifying four characteristic values of the desired cooling cycle, two pressures and two temperatures, as described in \autoref{sec:background}.

Point 1 (refer, again, to \autoref{fig:ph diagram general}) is determined by the low pressure and the superheating temperature. Initially, point $1'$ is located (on the saturation curve at the liquid side for the given pressure), and the corresponding temperature is identified from the iso-thermal curves (red lines in \autoref{fig:ph diagram general}). Subsequently, the specified amount of superheating temperature is added, and the position of the resulting temperature with the specified pressure constitutes point $1$. Interpolating from the iso-curves is not straightforward due to their non-linear gradient. We utilize the CoolProp library \cite{CoolPropLibrary} to obtain the desired points.

The cycle then follows the iso-entropy line (ideally) until it reaches the high pressure. In practice, we account for the compressor efficiency and adjust point $2$ accordingly. Subsequently, the cycle proceeds towards point $3'$, following the high pressure, and point $3$ is computed similarly to the computation of point $1$.  Finally, point $4$ shares the same enthalpy value as point $3$ and has the low pressure. The user also has the option to move the points using the mouse, and all of the aforementioned constraints will be enforced. Without automatic guidance, specifying point coordinates directly for the desired cooling cycle would be rather impractical. During our exploration sessions, engineers consistently entered numeric values for the four parameters, finding it more efficient than using mouse selections.

\begin{figure}[t!]
    \centering
    \includegraphics[width=.6
    \linewidth]{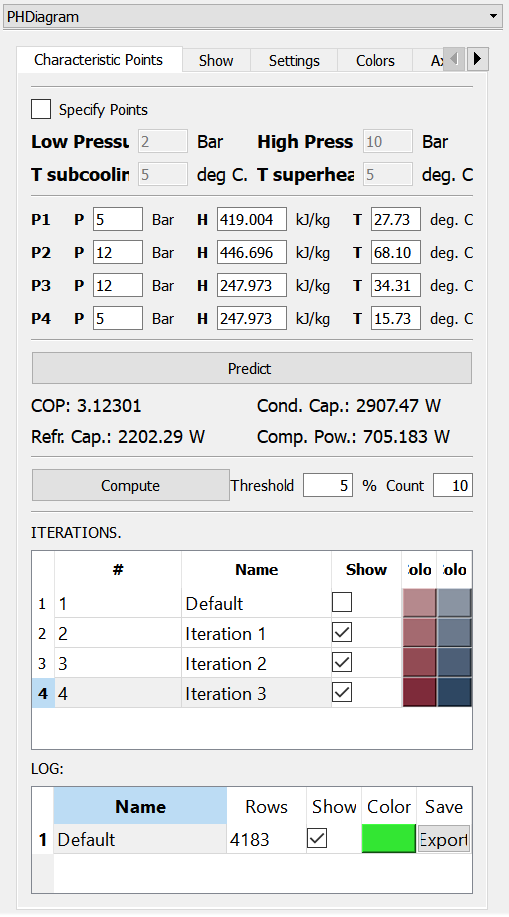}
    \caption{The control used to specify desired values (four values in the top), to provide feedback on aggregated data (the second block of values), to trigger ensemble refinements, and to keep track of iterations and  findings. }
    \label{fig:specify}
\end{figure}

The GUI elements for point specification are depicted in \autoref{fig:specify}. Once the four parameters are entered, pressure, enthalpy, and temperature are computed for each point. When a prediction by DL model is requested (R4), a new iteration is created. The corresponding control parameters are predicted along with the four values crucial for assessing the quality of the selected points: COP (coefficient of performance), condenser and refrigerant capacity, as well as compressor power. These four values are displayed in the GUI controls (\autoref{fig:specify}). Refinements for an iteration can also be triggered from the same control~(R5).

\paragraph{Providing Dual Perspective}

The p-h diagram explicitly depicts pressure and specific enthalpy, while temperature, quality, and entropy for each characteristic point can be inferred from nonlinear iso-lines. These values are utilized to compute various aggregated data. However, interpreting values based on these non-linear iso-lines is challenging. To simplify analysis, we provide a dual representation of the p-h diagram using four scatterplots (R3). Each scatterplot shows additional aggregated values closely related to its characteristic point, reflecting its position in the cycle. To aid association with the points, the scatterplots mirror the arrangement of the points themselves. We experimented with two alignment strategies: grouping the scatterplots together on one side of the p-h diagram or distributing them across two opposite sides (left-right or top-bottom) to align with the original points. The latter arrangement, with two plots on each side, was unanimously preferred by domain experts. \autoref{fig:teaser} illustrates the favored arrangement. The points in the scatterplots are colored using the same colors as those in the p-h diagram. Additionally, to aid in identifying corresponding cases, predicted and computed cases for the same parameter setting are connected by a line, assisting in relating them to the corresponding points in the p-h diagram.

\subsubsection{Additional Views and Integrated System}

%\autoref{fig:teaser} displays a screenshot of an analysis session. 
Besides the interactive p-h diagram, we have integrated additional views into a coordinated multiple view setup: the parallel coordinates view, the box plots view, and the details view. The details view presents detailed information for a selected subset in a tabular format. This view can be used to inspect individual runs after a drill-down or to export the values for use in other applications. The additional views provide information on control parameters, as well as different perspectives and statistics on some output values (R1). Composite brushing, which combines individual brushes across various views using Boolean operators, is supported.
The chosen color maps are used consistently across all views. Also the visibility state of an iteration, chosen in the p-h diagram, is propagated to all other views, ensuring a consistent perspective on the data. This makes it easy to link cases and iterations across different views (R6).

\paragraph{Parallel Coordinates}

We utilize parallel coordinates to depict the simulation control parameters (\autoref{fig:teaser} top right). They provide a proven method for displaying a multidimensional parameter space, and our domain expert is familiar with them. Therefore, they were the natural choice for visualizing the five control parameters. Still, some adaptations were needed. We use the parallel coordinates to show the initial ensemble, the values predicted by the DL model, and the refinements by the simulation solver. Each of these lines have different settings in order to be easily distinguished. In addition, the rendering order is fixed, so that the initial runs are rendered first, the refinements then, and the predicted values on top. 

\paragraph{Box Plots}

Seeing all computed values in scatter plots provides a good overview of distributions but lacks additional quantitative details of the output parameters (R7). To quickly compare different iterations, the box plots depict basic statistics for each one ( \autoref{fig:teaser} bottom right). This provides an informative overview of the development of basic descriptive statistical values across iterations. Besides the statistics for each iteration, the box plots also show the overall statistics of the initial ensemble, which serves as context. Again, the rendering order ensures maximum visibility.

\subsubsection{Visual Settings.}

%TODo: fix colors and references, check text, KM
We follow an iterative workflow, resulting in different outcomes for each iteration. Alongside the initially computed points, we have predicted parameters and computed cases based on these predictions, as well as several refinements at each iteration. To visualize these variations, we utilize point size and color, distinguishing between different cases and iterations (R6).

\begin{figure*}[t!]
    \centering
   \includegraphics[width=\linewidth]{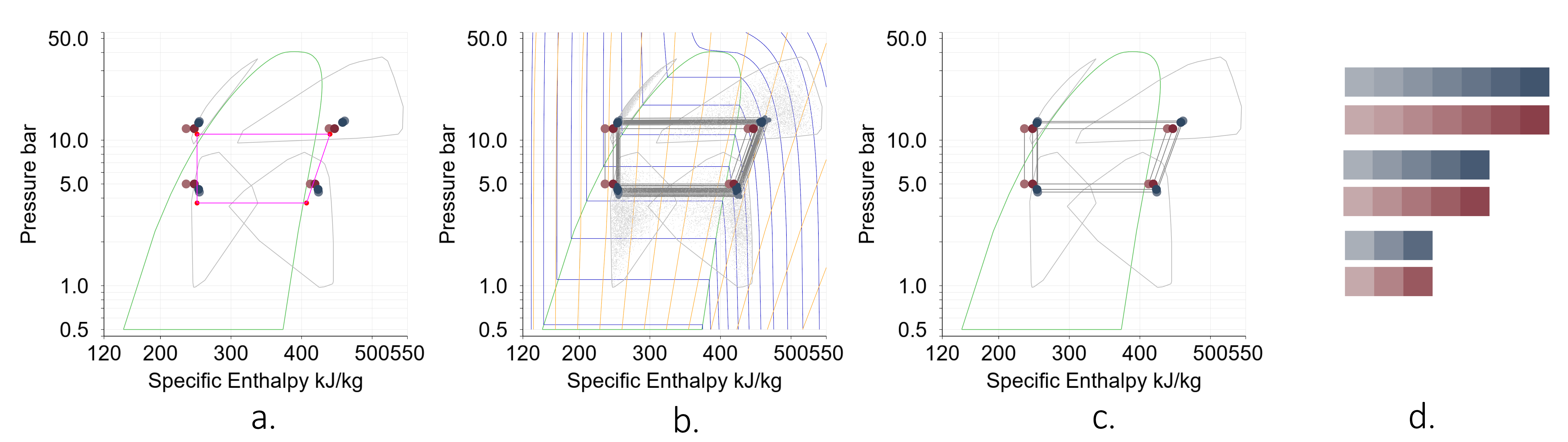}
    \caption{Interactive p-h diagram showing different information at different workflow stages. \textbf{a.} When specifying desired points (bright red), the reference curve (green), convex hulls of the initial points (gray), and previous iterations of interest are shown (red and blue). \textbf{b.} A multitude of information can be shown: quadrilaterals of cooling cycles for simulation runs (gray), initial points (gray), iso-thermal lines (blue), iso-enthorpic lines (orange). Showing all of the layers at once results in a cluttered view. \textbf{c.} The engineer focuses on two iterations and hides many layers. Convex hulls and the reference curve provide context. \textbf{d.} We use seven-level color maps. The current iteration is mapped consistently to the darkest shade. The step in the color map is adjusted based on the number of iterations to enhance the contrast between iterations. }
    \label{fig:points}
\end{figure*}

Given that analysis sessions typically involve a limited number of iterations, we opted for a sequential color map with seven levels for both predicted and computed cases (one map for each). The darkest color represents the most recent iteration, while progressively lighter shades indicate earlier iterations. When there are only a few iterations, we avoid using neighboring color map levels to enhance the distinction between iterations. If more than seven iterations are present, older ones share the same color.
\autoref{fig:points}d displays two color maps, red and blue, alongside colors for three and five iterations. %, generated using ColorBrewer 2.0 \cite{colorbrewer}.

To facilitate the association of predicted and simulated values, we connect corresponding points with a line. Furthermore, additional cases computed by the simulation solver are depicted using the simulation color map, albeit with a smaller point size to differentiate them from cases computed using parameters predicted by the DL model.

\autoref{fig:points}b showcases a detail of the p-h diagram, depicting initial points (gray), predicted points (red), and simulated points (blue) across two iterations. The convex hull of the initial points (gray), along with iso-thermal lines (blue) and iso-entropy lines (orange), are also depicted. Additionally, gray quadrilaterals represent individual cooling cycles by connecting characteristic points of a single run. Although the display is cluttered with all this information, each layer is essential at specific workflow stages. \autoref{fig:points}a shows the configuration used for specifying desired values, while \autoref{fig:points}c shows a configuration when an engineer examines two iterations and focuses on the characteristic points only. The reference curve (green) and the convex hulls of %initial point values corresponding to a plausible range are included for context.
initial point values corresponding to plausible parameters are included for context.

\subsubsection{Iterations and Findings Bookkeeping}

% \begin{figure}[t!]
%     \centering
%     \includegraphics[width=\linewidth]{Figures/Bookkeeping.png}
%     \caption{missing caption}
%     \label{fig:bookkeeping}
% \end{figure}

To facilitate an overview of iterations and manage them effectively, we provide a GUI element as depicted in the lower part of the \autoref{fig:specify}. For each iteration, it offers the option to change the name, displays information on the colors used for predicted and simulated cases, and includes a checkbox to toggle the visibility of that iteration (R6).

In line with the strategy proposed by Yang et al.~\cite{Ward_NUgget} and in response to users' need to keep track of significant findings, we also offer a logging feature for certain cases. When users isolate a single case or a small set of related cases using interactive drill-down, they can be stored in a log of findings. Users can customize the name, toggle the visibility, adjust the color, or export cases of a finding to a CSV file for reporting or further analysis in a different tool.

\subsection{The Deep Learning Component}

The DL component is used to estimate an inverse simulation model based on the available simulation runs. 
In reverse to the simulation model, the DL model expects four characteristic points as input, and provides the corresponding control parameters, five in our case, for the simulation model as outputs (R4).

%In our case, the goal of our deep learning model was to process four characteristic points from the p-h diagram and output six parameters of the AC system simulation model. 
 
In the Batch Learning Process of our study, initially, simulation data generated from the AC system runs were % meticulously 
collected and compiled into a comprehensive dataset. This dataset was crucial for the subsequent batch learning phase, where it was processed by a specifically designed DL module dedicated to batch processing. Embracing a traditional yet strategic approach, the DL model, identified to tackle the inherent non-linear regression problem of our study, was initially approached with simplicity, setting a baseline for further enhancement.
% \begin{figure}[t!]
%     \includegraphics[width=\linewidth]{Figures/MSE.png}
%     \caption{Model depth comparison.}
%     \label{fig:chartMSE}
% \end{figure}

\begin{figure}[t!]
\centering
    \includegraphics[width=.8\linewidth]{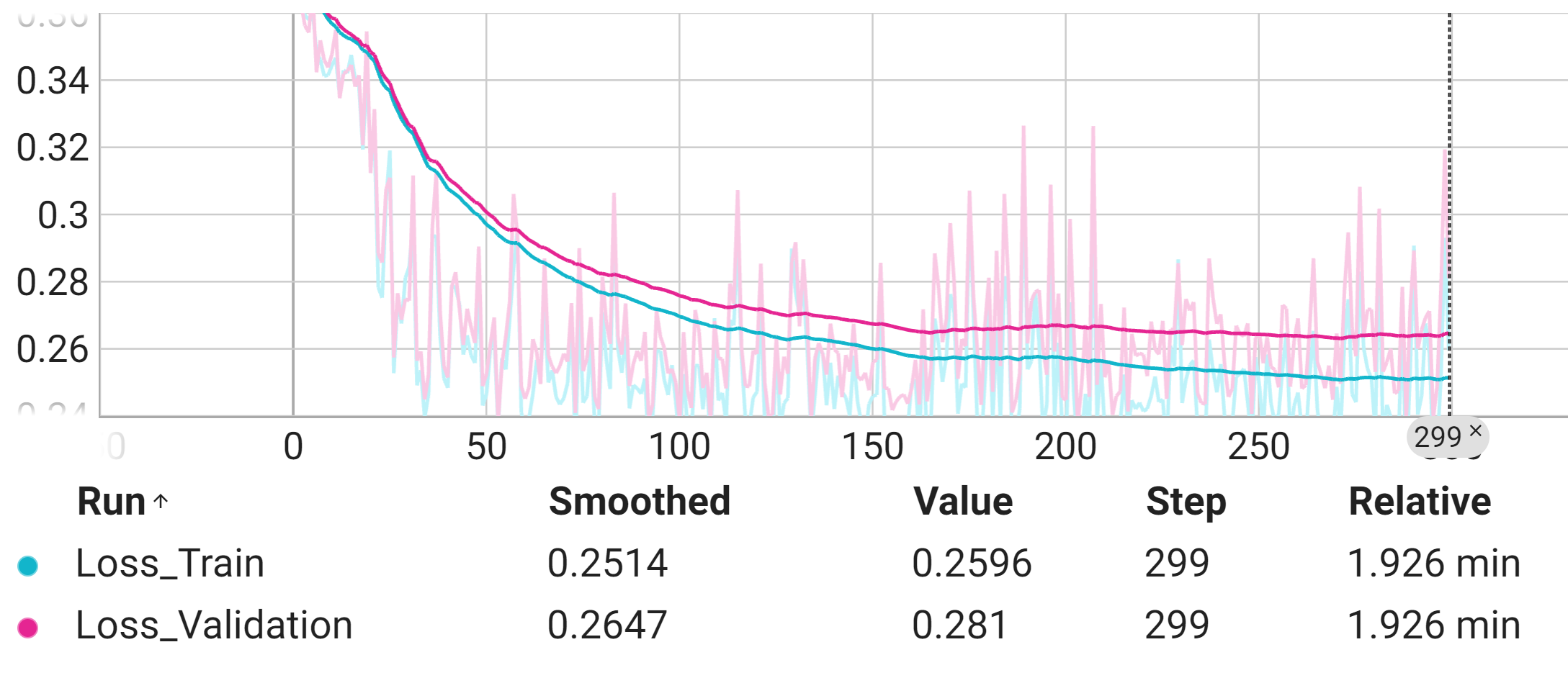}
    \caption{Train and validation loss.}
    \label{fig:chartLoss}
\end{figure}
 
We implemented a deep neural network comprised entirely of fully connected layers. Our analysis revealed that increasing model depth beyond three hidden layers did not enhance accuracy but instead increased complexity, parameters, and training time. Therefore, we opted for a model architecture featuring three hidden layers, each employing a tangent hyperbolic (Tanh) activation function, except for the output layer. Hyperparameter optimization was conducted using an exhaustive grid and basin search methodology, ensuring the model's robustness and efficiency.
 
The training of this model was a meticulous process, spanning 300 epochs, employing the Mean Squared Error Loss function, and leveraging the AdamW optimizer. This was complemented by precise adjustments to learning rate and weight decay, reflecting our commitment to refined model tuning. The fruit of this labor was evident in the model’s performance, which registered an impressive average Mean Squared Error (MSE) of 0.25 on the test dataset—a testament to the model’s robustness and the efficacy of our approach.

The dataset, an aggregation from 4182 simulation runs, was strategically partitioned, allocating 80\% for training, and 10\% for validation and 10\% for testing. The structure enabled batch learning, updating model weights iteratively after processing each batch. This approach ensured continuous refinement across epochs and consistent evaluation, validating effective learning.
%It exemplifies our research's integration of theory and practice, advancing simulation-driven machine learning.
This model architecture was supported by empirical findings detailed in \autoref{fig:chartLoss} and showcasing comparisons of  training/validation loss. 
 
Our study employs an Online Learning Enhancement strategy,
enabling continuous model adaptation through incremental learning. This dynamic interaction ensures the model's relevance and accuracy over time, reflecting the evolving conditions of the simulated AC system (see workflow in \autoref{fig:chartDLModel}).
%alongside batch learning. 
%featuring an interactive visualization module for real-time data acquisition and analysis. This setup integrates seamlessly with the `Deep Learning (online processing API)' module, enabling continuous model adaptation through incremental learning. This dynamic interaction ensures the model's relevance and accuracy over time, reflecting the evolving conditions of the simulated AC system (see workflow in \autoref{fig:chartDLModel}).
 
Further, our Iterative Feedback and Model Updating process establishes a robust feedback loop. Here, predictions from the deep learning component guide control parameters in the simulator, fostering a cycle of prediction, simulation, and data re-acquisition. This iterative approach, integrating batch and online learning methodologies, continuously refines the `Updated Model', enhancing its precision and adaptability to real-world data dynamics.
 
Overall, our methodology harmonizes simulation and learning components, leveraging historical data for foundational learning while adapting to new insights. This integrated approach ensures the model's accuracy and flexibility in addressing the complex and dynamic demands of AC system simulation.

\subsection{The Simulation Component}

\begin{figure}[t!]
    \includegraphics[width=\linewidth]{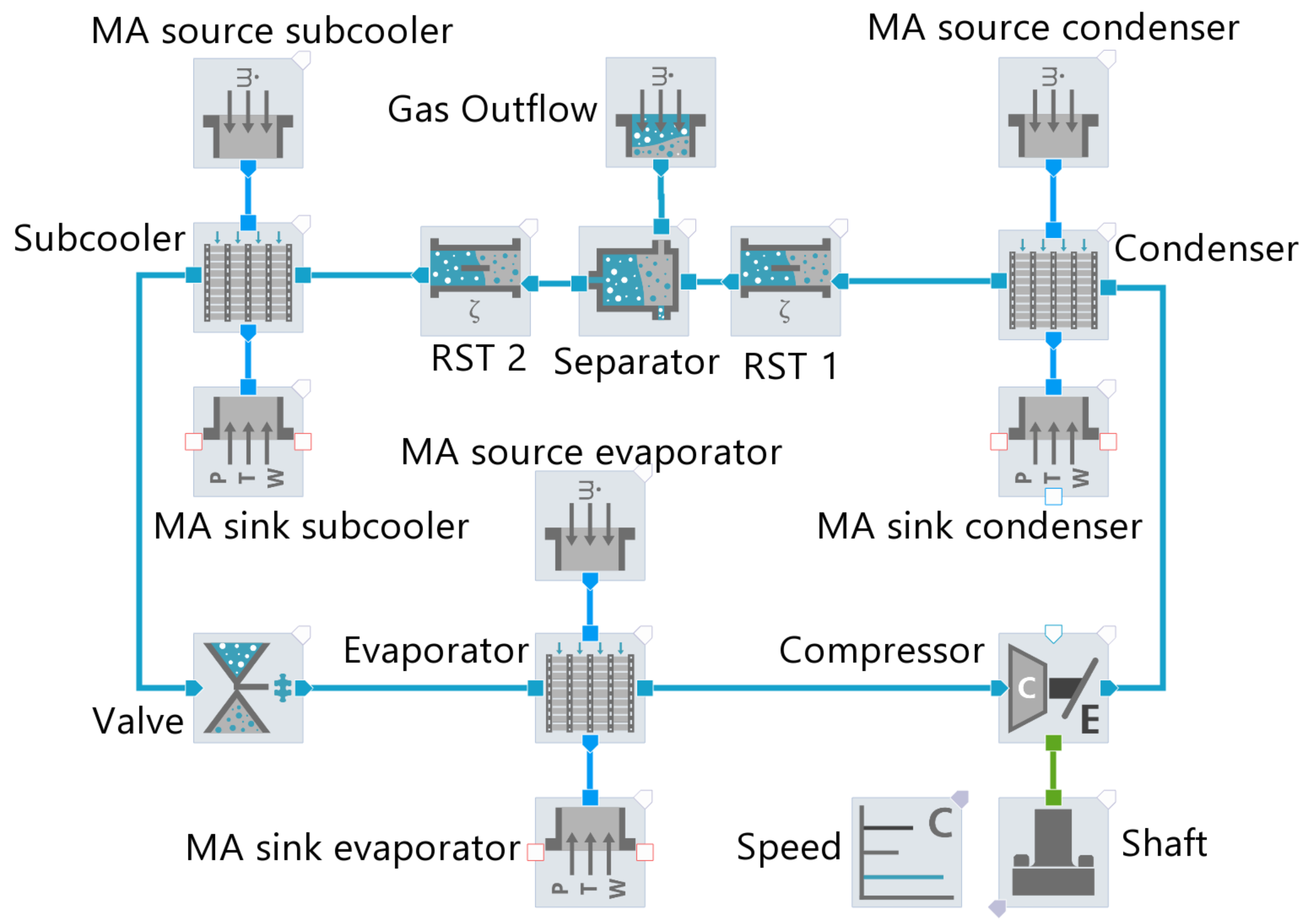}
    \caption{The simulation model of an AC system.}
    \label{fig:simModel}
\end{figure}

We use the multidisciplinary vehicle system simulation tool AVL CRUISE\texttrademark M, which is commercially available and can simulate AC systems. 
The user defines the simulation model using basic building blocks (\autoref{fig:simModel}), sets the simulation control parameters, and triggers the numerical simulation.
The simulation computes the four characteristic points in the p-h diagram, and a lot of additional output values.

In our model the condenser, subcooler and evaporator are vapor, liquid and equilibrium (VLE) multi port extruded tubes (MPET) heat exchanger components.
%VLE stands for vapor, liquid and equilibrium.
%In this kind of circuits two phase flows are simulated.
The MPET heat exchanger is a crucial component of a VLE circuit, used to simulate the heat transfer between moist air and a VLE fluid as well as the pressure-drop phenomena through the heat exchanger of a user-defined geometry.
The simulation solver for such complex thermodynamic cycles are characterized by their non-linear behavior. 
%These models can only be linearised for a small range of validity. 
The physical system models are often solved numerically in form of an ordinary differential equation system or differential algebraic equation system.

%The simulation solver for complex thermodynamic physical systems, in this case, a thermodynamic cycles, are characterised by their non-linear behaviour. These models can only be linearised for a small range of validity. Therefore the physical system models, instead of black box models, are often solved numerically in form of an ordinary differential equation system (ODE) or differential algebraic equation system (DAE). The solving of a physical model of such a thermodynamic system is far more complex than solving a black box model.
%More sophisticated than the description of pure fluids for simulation is the description of mixtures. Although interpolation methods could be applied to describe the whole fluid region, this is not done due to the tremendous amount of data required. Often only the calculation of the VLE is optimised, whereas the one-phase region is still calculated using empirical multiparameter equations of state (EoS).

It is impossible to mathematically invert such a simulation model, i.e., it is not possible to specify the four desired characteristic points and to get the corresponding simulation control parameters. %that would generate them.
Due to the complexities of some elements in the model the computation time for a single simulation run is approximately ten seconds.
Hence the computation for our initial ensemble consisting of 5000 runs took about 14 hours.
Since the solver did not produce valid results for all runs we end up with 4182 runs in the initial ensemble.

% \subsection{Typical Visualizations}

% We have encountered several typical visualizations which clearly indicate certain data characteristics. 
% We briefly describe them now.

% \paragraph{Out of range parameters}

% \paragraph{Large deviation}

% \paragraph{Threshold too small}

%%%%%%%%%%%%%%%%%%%%%%%%%%%%%%%%%%%%%%%%
%%%%%%
%%%%%%  Use Case
%%%%%%
%%%%%%%%%%%%%%%%%%%%%%%%%%%%%%%%%%%%%%%%

\section{Use Case}

%In the process of designing a cooling system, engineers are faced with the challenge of creating an efficient and reliable system that meets the required specifications. There are several approaches that can be taken to tackle this task. One approach is to build a physical prototype of the cooling system and test it in various environments to assess its performance. While this method provides valuable real-world data, it can be prohibitively expensive and time-consuming. Alternatively, engineers can opt to develop a simulation model of the cooling system. By carefully tuning the parameters of the model, it can be made to closely approximate the behavior of the real system. This approach offers the advantage of being cost-effective and allowing for rapid iteration and experimentation.

In our use case, we begin with a fine-tuned simulation model that we use to explore diverse scenarios and observe the system's behavior under various conditions. This yields valuable insights into the system's performance and highlights potential areas for improvement.
However, it is important to note that interpreting the results of simulation models is not a straightforward task. The behavior of the AC system is influenced by a multitude of factors, and the interactions between these factors can be highly complex. The engineers must have a deep understanding of the system's behavior in order to accurately interpret the results of simulations.
%The newly proposed approach is a valuable support in this undertaking.

We are dealing with an AC system designed to cool the cabin of an electric vehicle, where the cabin temperature is set to fluctuate between 20 and 50 degrees Celsius, with an external temperature of 30 degrees Celsius, representative of a typical summer day.

Our objective is to delve into the system and analyze the values requiring control parameter adjustments to achieve the desired functionality. The challenge involves pushing the system beyond current boundaries, exploring the potential for expansion while facilitating straightforward upgrades to avoid extensive interventions. 
%Future development directions are also of interest, with engineers keen on avoiding the creation of entirely new AC hardware to save time and resources.

We seek parameter settings that maintain stability within the parameter space. While a highly efficient but overly sensitive solution is undesirable due to its susceptibility to minor parameter changes, it's crucial to ensure that the parameters remain feasible and adaptable to real-world fluctuations during system operation.
We use the simulation model as shown in \autoref{fig:simModel} and we vary five parameters to create the initial ensemble. We set the physically plausible boundaries and use Latin Hypercube Sampling~\cite{LatinCube} to create individual cases. The five varied control parameters are:
\begin{enumerate}
\item $N\_pump$ -- speed of the compressor pump. 
\item $MF\_air\_cond$ -- mass flow of ambient air through which heat is expelled within the condenser. %A fraction of the mass flow is used in the subcooler. %COMMENT Mario, we could stress out that this mass flow of the air whichj was taken from ambient to the condenser and then it is expeled to ambient agaian.
\item $T\_air\_cabin$ -- temperature of cabin air to be cooled in evaporator.
\item $MF\_air\_evap$ -- mass flow of cabin air cooled in the evaporator. %mario: we can say this is the mass flow of the cool air that comes to the cabin
\item $A\_eff\_valve$ -- effective flow area of the valve, used to calculate the actual flow rate of the valve orifice.
\end{enumerate}

\noindent In addition to the four characteristic points, the simulation solver also computes the following output values relevant for the analysis:
%which will be considered during the analysis:
\begin{enumerate}
    \item $COP$ -- coefficient of performance, indicates how effective an AC is at transferring heat vs. power consumed by the compressor. 
    %It is computed as $COP = (h_1-h_4)/(h_2-h_1)$
    \item $T_{subcooling}$ -- subcooling temperature (see \autoref{sec:background}).    
    \item $T_{superheating}$ -- superheating temperature (see \autoref{sec:background}).     
    \item $dh_e$ -- condenser capacity, the ability of the condenser to transfer heat from the hot vapor refrigerant to the condensing medium.
    \item $dh_c$ -- refrigerant capacity, the ability of the evaporator to transfer heat from the hot outside media to the refrigerant.
    \item $W$ -- electrical power consumption of the compressor.
\end{enumerate}

\noindent A previous exploration of a scenario confirmed that the model functions properly 
for the following values of pressures and temperatures: $P_{low} = 2.5\;bar$; $P_{high} = 12\;bar$; $T_{subcooling} =  7^{\circ} C$, $T_{superheating} =  8^{\circ} C$. In the current scenario, which we describe now, the goal is then to increase the coefficient of performance (COP).

\begin{figure}[t!]
    \includegraphics[width=\linewidth]{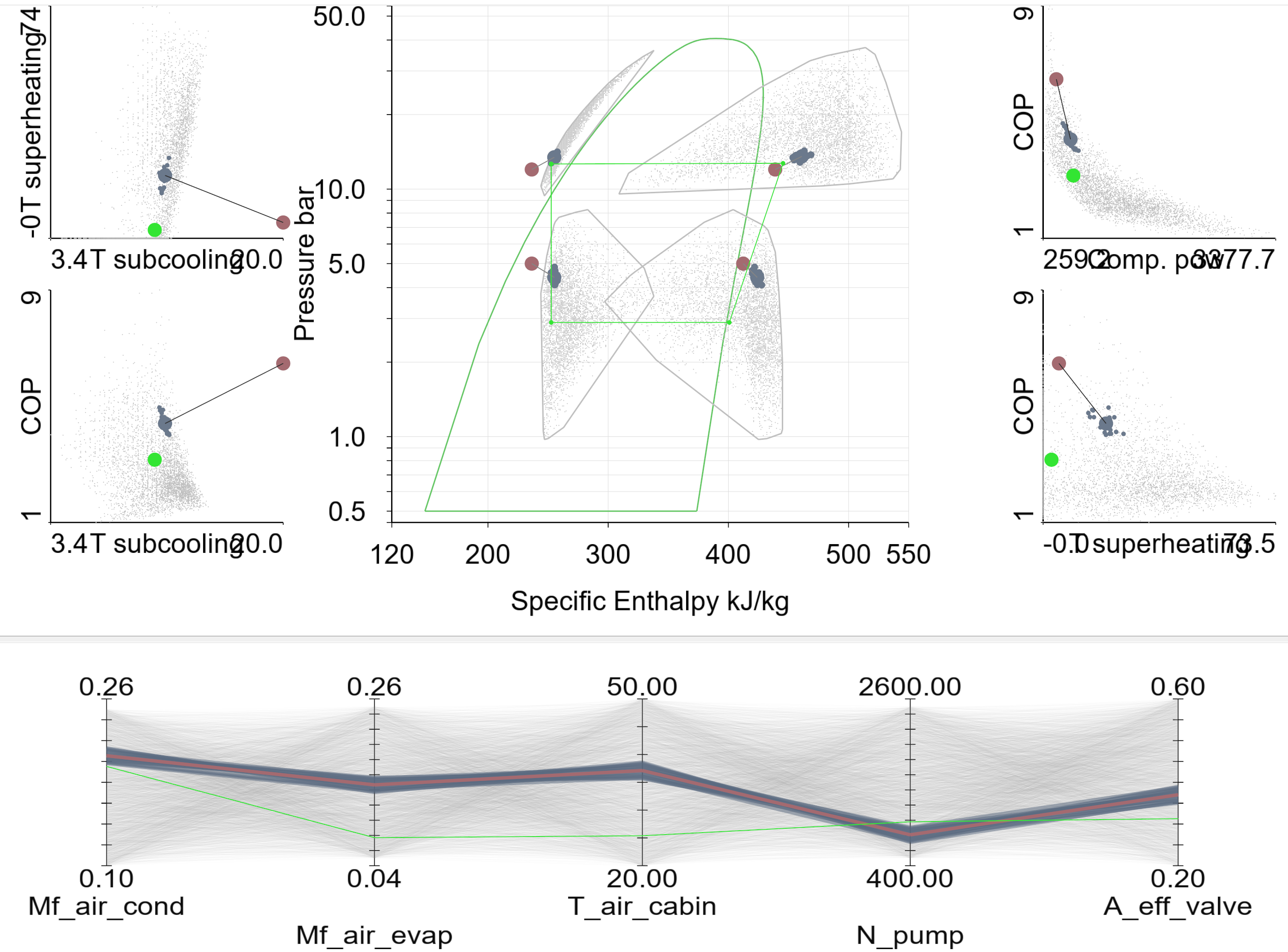}
    \caption{First iteration: The user specified values are shown in red. They lie outside the plausible range. The DL suggests parameters which result in the blue points when simulated. They do not match the desired target, but they are plausible. The light green point in the scatterplots shows the starting scenario. We achieved the goal of increasing COP.}
    \label{fig:usecase1}
\end{figure}

\begin{figure}[t!]
\centering
    \includegraphics[width=.8\linewidth]{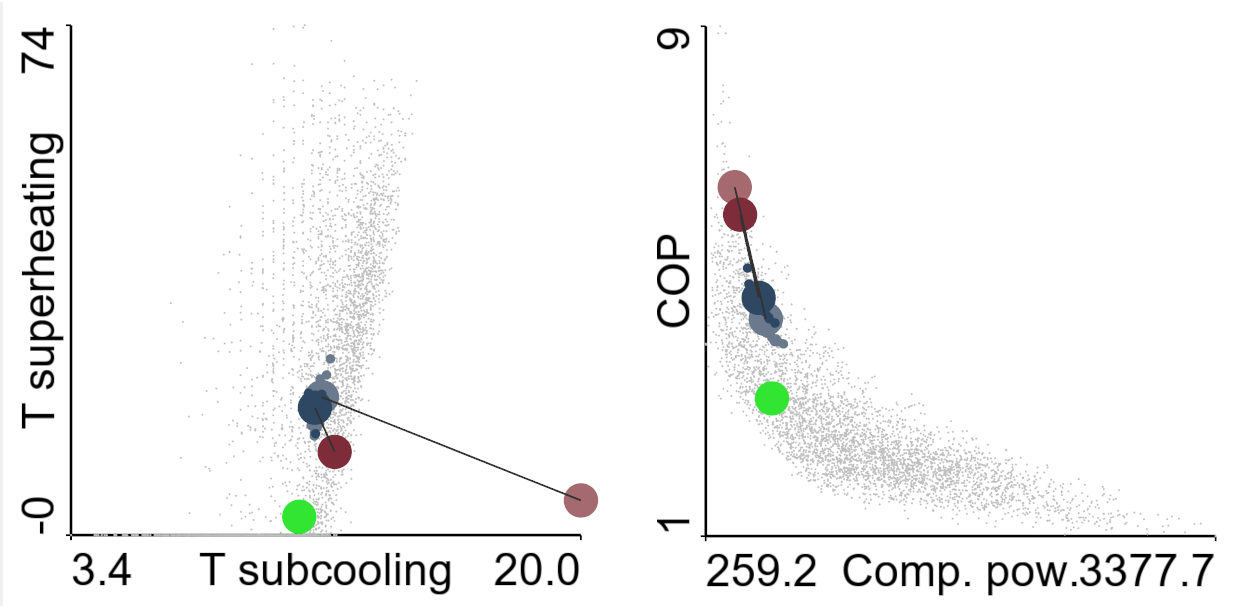}
    \caption{\textbf{Left:} Temperatures for the predicted (red) and simulated values (blue). The first desired subcooling temperature (light red point)  was set unrealistically high, and the DL model pushed it back into plausible range.
    \textbf{Right:} The COP values for the simulated refinements are significantly higher than for the starting scenario (light green point).}
    \label{fig:usecase2}
\end{figure}

% \begin{figure}[t!]
%     \includegraphics[width=.7\linewidth]{Figures/UseCase_iteration3-SP.png}
%     \caption{The COP values for the simulated refinements are significantly higher than default value (shown as the light green point). }
%     \label{fig:usecase3}
% \end{figure}

\paragraph{Iteration 1.}
The desired value for $P_{low}$ is increased to $5\;bar$, while $P_{high}$ remains at $12\;bar$. To enhance the refrigerant capacity, the subcooling temperature is elevated to $20^{\circ} C$, and we set the superheating temperature to $5^{\circ} C$. We do not lower it below $5^{\circ} C$ to mitigate the risk of harmful compressor temperatures. %COMMENT MJ: IT could be higher. It is important that it is not too low.

A prediction of the control parameters is requested from the DL model, and the results are immediately available. Subsequently, we commence the computation of twenty runs with the simulation solver, varying parameters by up to 5\% relative to the predicted values. The results are explored in the visualization.

We use the three main views, the augmented p-h diagram, parallel coordinates, and boxplots, as shown in \autoref{fig:teaser}. The scatterplots around the p-h diagram show data relevant for the use case (note that COP is shown in several plots). The parallel coordinates show the control parameters, and the box plots show the three aggregated values which indicate how much energy flows through the system. It is necessary to immediately see if the energy flow is too high. 

The parallel coordinates in the \autoref{fig:usecase1} show that the control parameters for the first iteration fall quite central within the original ranges (the DL model does not generate solutions outside plausible boundaries). 
In the p-h diagram we can see that requested point 3 (red point) is quite far from the plausible area (gray convex hull). 
The simulated refinements (blue points), however, fall within the original available area. The connected line between desired and simulated point indicates this difference.
We proceed to examine other characteristics of the simulated cases, particularly focusing on $COP$ and $T_{superheating}$. Notably, the desired $T_{superheating}$ lies significantly outside the plausible area. However, the simulated cases maintain $T_{superheating}$ within the acceptable range, albeit with several different values observed. Thus, we anticipate finding a suitable solution.

The $COP$ values are satisfactory, prompting us to proceed to the next iteration. We show the starting scenario using bright green color. Computed cases have a higher COP values.
We opt to reduce $T_{subcooling}$ in the subsequent iteration, aiming to achieve an optimal solution.

\paragraph{Iteration 2.}

We adjust $T_{subcooling}$ to align it closer to real values, setting it to $12^{\circ} C$.
The results indicate that we remain within a similar range as in the initial iteration, with $T_{subcooling}$ values around $11^{\circ} C$.
Consequently, we decide to retain this value and proceed to explore other parameters.
The scatterplot in \autoref{fig:usecase2}~left shows the wanted values of the temperatures in red for the first two iterations, and corresponding simulated values in blue.
The first subcooling temperature was excessively high, rendering it unattainable through simulation.
The second desired value proved to be more realistic, although it also required correction.
The superheating temperature was set closer to the computed values, the shift in the y-direction is much smaller. 

Since we aim to maintain consistent pressures (which correspond to ambient temperatures), we refrain from altering them.
Instead, we focus on increasing $T_{superheating}$ to observe its effects.
However, caution is warranted to avoid excessively high values or those below $0^{\circ} C$, as such extremes could potentially damage the compressor.

\paragraph{Iteration 3.}

We adjust  $T_{superheating}$ to $12^{\circ} C$ and repeat the procedure.
After computing the simulation runs, we obtain a potential configuration for our simulation model.
The coefficient of performance (COP) shows a reasonable increase compared to the original scenario (\autoref{fig:usecase2}~right).
Additional control over cabin temperature indicates that the system can maintain the desired characteristics even with temperature variations.
Cases in the vicinity of the selected case exhibit similar temperatures (see, for example, the parallel coordinates in \autoref{fig:teaser}).
Ensuring close proximity of cabin temperatures avoids the need for large adjustments in the system settings for minor temperature changes.

This brief use case demonstrates our ability to significantly enhance the COP in just three iterations. We swiftly corrected the initial setting and pinpointed the promising area for improvement. The fine-tuning process, coupled with sensitivity analysis of the chosen solution, proved highly efficient. Achieving such an increase in COP through conventional methods would typically entail extensive trial and error, resulting in a more laborious workflow. The exploration and tuning continues by examining further scenarios in the same manner.

\section{Discussion}

Designing a visual analysis system for domain experts poses significant challenges to both the visualization designers and the domain experts. The IDoE approach emerged from a participatory design and development process, which was lengthy and involved numerous discussions, clarifications, and iterations of the design, including various color scales, view configurations, and more. Reflecting on the process, we cannot overstate the importance of repeatedly discussing visual analytics principles and learning the basic principles of the domain.

In our case, we were fortunate that the domain expert already had experience with interactive visual analysis. On the other hand, we had to learn extensively about AC systems. Familiarity with the domain is crucial for success. It became evident to us that the p-h diagram had to be the central view. Given our five-dimensional parameter space, we suggested using parallel coordinates. Interestingly, with visualization tools being more widely used in the past decade or two, there are more experts familiar with parallel coordinates. Our main collaborator was aware of them, and his colleagues, to whom we presented the system, quickly understood them. Despite some drawbacks, we could not find a better way to depict the 5D parameter space.

The decision to use a consistent color scale across all views proved to be correct and very important. We experimented with many color scales and, following also the reviewers' suggestions, finally selected red and blue colors.

The time of domain experts is usually very valuable as they are engaged in various projects within their companies. Finding an expert willing to engage in developing a new approach was crucial. To maintain the expert’s enthusiasm and prevent them from leaving the project early, the proposed system had to offer significant advantages compared to the current workflow. After initial discussions with the domain expert, it was essential to present a prototype that captured the expert’s attention and clearly demonstrated its usefulness. Of course, the first prototype was far from a fully operational system, but it had to clearly show its potential benefits.

General design guidelines suggest not offering too much flexibility. John Maeda~\cite{Meada_TheLawsofSimplicity}, for example, explores the concept of simplicity in design and how it can lead to better user experiences. One of the laws discussed is ``Reduce'', which suggests that removing unnecessary complexity can improve usability. At the same time, Jenny Preece et al.~\cite{Price_InteractionDesign}, e.g., discuss the importance of flexibility in interfaces to support diverse user needs and workflows, especially for expert users. Since we cater to experts, and different requirements may arise depending on the current workflow state, we offer a highly flexible system where individual layers can be easily toggled on or off. 
We organize all customization options into different semantic tabs (e.g., Show tab, Characteristic Points tab, etc.) to facilitate quick access to individual settings. 
After a brief customization period, experts were satisfied with the organization, finding it easy to locate the required settings promptly.

When reflecting back on the feedback of the domain users, one of the most appreciated aspects of the method is the tight integration of all three components: deep learning, simulation, and visualization. Tuning an AC system, as described in the paper, would otherwise necessitate multiple separate tools and cumbersome data exchange between them. 
%The integrated system proves far more efficient to use.
The integration of various aspects within a single workflow, along with diagrams, that are familiar to the domain experts and are augmented for enhanced understanding, notably boosts the efficiency of exploration and prediction processes, as reported by the domain experts.
They further attested that the IDoE approach accelerates their cooling system design tasks by at least an order of magnitude. An analysis that took several days can now be accomplished in few hours, at most.
% We did not formally measure the speed-up, but engineers stated that the new IDoE approach accelerates their cooling system design tasks significantly. 

% The IDoE approach has been designed for AC Systems, but it can certainly be applied to other problems where inverting a simulation model is not possible. The limitations of the newly proposed approach include the number of control parameters and the speed of simulation. Using parallel coordinates, the reasonable limit on the number of control parameters is about a dozen. Additionally, to generate refinements, the simulation needs to be relatively fast. In our case, the engineers were willing to wait a few minutes for the refinements. For longer simulations, the wait might be excessive. While parallelization could speed up the process, each individual run still needs to be completed quickly. The definition of "quickly" can vary depending on the domain, as users in some domains are accustomed to waiting longer for results.

The IDoE approach, designed for AC Systems, can also be applied to other problems where inverting a simulation model is not possible. Limitations include the number of control parameters and the speed of simulation. Using parallel coordinates, the practical limit is about a dozen control parameters. Simulations need to be relatively fast to generate refinements; our engineers were willing to wait a few minutes for a refinement with twenty runs. Longer waits may be impractical, though parallelization can help. The definition of "quickly" varies by domain, as some users are accustomed to longer waits.
%%%%%%%%%%%%%%%%%%%%%%%%%%%%%%%%%%%%%%%%
%%%%%%
%%%%%%  Implementation
%%%%%%
%%%%%%%%%%%%%%%%%%%%%%%%%%%%%%%%%%%%%%%%

%\section{Implementation}

%The visualization component is implemented using C++ and QT and is a research prototype.
%The simulation component uses Python and C++ and is a commercially available software.
%The DL component is implemented in Python.
%The integration of simulation, DL, and visualization is achieved by implementing a network interface for all three components.
%A central dispatch server written in Python handles the messages and routes them to the target specified in the message.
%This architecture allows to send requests and responses between all components in a very flexible manner.

%%%%%%%%%%%%%%%%%%%%%%%%%%%%%%%%%%%%%%%%
%%%%%%
%%%%%%  Conclusion
%%%%%%
%%%%%%%%%%%%%%%%%%%%%%%%%%%%%%%%%%%%%%%%

\section{Conclusion}

Whenever a simulation model cannot be inverted or inversion proves overly complex,  interactive design of experiments (IDoE) emerges as a potent approach for exploration and analysis. 
This method ensures that additional simulation runs are strategically computed in areas of the control parameter space likely to yield desired outputs.
The general concept is applicable across various domains.
%particularly beneficial for simulation models that cannot be easily inverted.

We acknowledge the complexity of the user interface and that its usability could undoubtedly be improved. However, mechanical engineers who utilized the system quickly became acquainted with it and utilized it seamlessly. Their familiarity with complex simulation software and their expectation of a learning curve likely contributed to its favorable reception.

Looking ahead, future research should capitalize on advancements such as deep learning to further expedite simulation solver speed and augment the tool's capabilities.
This includes incorporating additional constraints in the specification of the design goals in the DL prediction, while also expanding the interface support accordingly, 
enabling the exploration of scenarios outside of the default control parameters space,
and implementing an adaptive interface that flexibly adjusts to the user's needs. By continuously refining and expanding upon these advancements, we can empower experts to navigate the complexities of diverse engineering systems with enhanced efficacy and precision.

% Stress out importance of multidisciplinary team.

% -comment on necessity for flexibility and complex GUI as this is a tool for experts, and the need to show different aspect changes along the workflow and depending on tasks.

% -feedback from domain expert:
%     - much faster exploration than before
%     - ability to predict values for desired saves a lot of trial and error steps
%     - use of well known diagram is essential, augmenting it makes it even better
%     - integration f the three main components in a single workflow saves a lot of time
%     - It is impossible to exactly quantify speed up, but the new approach is certainly at least an order of magnitude faster.

%   Generalization, where applicable

%   Future work: solver speed up by DL,
%   - additional constraints in specification, not only desired values, but also settings of some of the control parameters
%   - Extend interface to support it
%   - Allow control parameters to go outside of the plausible areas to explore more extreme what if scenarios.
%   - Adaptive interface that controls flexibility

%%%%%%%%%%%%%%%%%%%%%%%%%%%%%%%%%%%%%%%%
%%%%%%
%%%%%%  acknowledgments
%%%%%%
%%%%%%%%%%%%%%%%%%%%%%%%%%%%%%%%%%%%%%%%

\clearpage
%% if specified like this the section will be ommitted in review mode
\acknowledgments{%
  %The authors would like to thank Damjan Rafaj for the first version of the DL model. 
  VRVis is funded by BMK, BMDW, Styria, SFG, Tyrol, and Vienna Business Agency in the scope of COMET -- Competence Centers for Excellent Technologies (879730), which is managed by FFG.  Parts of this work have been done in the context of CEDAS, i.e., the Center for Data Science at the University of Bergen, Norway.  
}

\def\UrlBreaks{\do\/\do-}
\bibliographystyle{abbrv-doi-hyperref}

\bibliography{CoolingPaper}

\end{document}